%% file: main.tex
\documentclass{topoclaw}
\begin{document}
\thispagestyle{firststyle}

\begin{center}
    \vskip 8mm
    {\color{topoblue}\hrule height 2pt}\par
    \vskip 5mm
    {\titlestyle\color{black}TopoClaw: A Human-Centric and Topology-Aware Agent Operating System \par}
    \vskip 5mm
    {\authorstyle
        Heyuan Huang\textsuperscript{*} \quad
        Yeyi Guan\textsuperscript{*} \quad
        Jihong Wang\textsuperscript{*} \quad
        Mingzhi Wang\textsuperscript{*} \par
        \vskip 1mm
        Jiamu Zhou \quad
        Xiangmou Qu \quad
        Jiaxin Yin \quad
        Xin Liao \quad
        Xingyu Lou\textsuperscript{\textdagger} \quad
        Jun Wang\par
    }
    \vskip 2mm
    {\affiliationstyle OPPO Research Institute \quad MadeAgents Team\par}
    \vskip 1.5mm
    {\fontsize{8.2}{9.5}\selectfont\color{topoblue}
        \texttt{\{huangheyuan2, guanyeyi, wangjihong, wangmingzhi, louxingyu\}@oppo.com}
    }
    \vskip 5mm
    {\color{topoblue}\hrule height 0.5pt}\par
    \vskip 8mm
\end{center}
\begingroup
\renewcommand{\thefootnote}{}
\footnotetext{\hspace{-1.8em}\textsuperscript{*}Equal contribution. \quad \textsuperscript{\textdagger}Corresponding author.}
\endgroup

\vskip 10mm

\begin{abox}
    \setlength{\parindent}{0pt}
    \setlength{\parskip}{0pt}

    \begin{center}
        {\abstractlabelstyle Abstract}\par
        \vskip 3mm
    \end{center}

    Large language models (LLMs) have turned AI assistants from conversational interfaces into autonomous reasoning engines that maintain context, invoke tools, and pursue long-horizon tasks. This shift has spurred the Agent Operating System (Agent OS) as a kernel-like layer for lifecycle management, memory, scheduling, and access control. Yet most practical designs remain \emph{agent-centric}: they treat the OS mainly as a runtime container on a single host that sustains internal reasoning and tool use. That framing leaves open how autonomous actions should integrate with the user's real workflows across a distributed, collaborative, and permission-sensitive digital life. TopoClaw is an open-source, \emph{human-centric}, topology-aware Agent OS that models the user's ecosystem as two coupled structures---a \emph{physical device topology} of heterogeneous surfaces and a \emph{social relationship topology} of shared spaces, teams, and delegated roles---and unifies device operation, messaging, and skills around accountable cross-boundary execution. Its architecture centers on three contributions: (1) \textbf{cross-device action placement}, decoupling intent from actuation and routing distributed actions across the device cluster according to hardware affordances and user context; (2) \textbf{cross-user identity attribution}, treating agents as socially situated ``Digital Twins'' that coordinate in multi-user spaces while preserving provenance, role-aware permissions, and a clear chain of human accountability; and (3) \textbf{cross-context authority governance}, pairing broad capability with distributed, context-aware policy enforcement across both physical and social trust boundaries so that proactive autonomy remains structurally bounded at the OS layer. This technical report presents TopoClaw as an engineering-oriented reference architecture and surveys its design principles, runtime, cross-device execution, collaboration mechanisms, security model, and deployment outlook.

    \vskip 6mm
    \parbox[c][20mm][c]{\linewidth}{
        \parbox{0.65\linewidth}{
            \textbf{GitHub:} \texttt{\url {https://github.com/MadeAgents/TopoClaw}} \\
            \textbf{Project Page:} \texttt{\url{https://madeagents.ai/news/topoclaw/}} \\
        }
        \hfill
        \raisebox{-10mm}{\includegraphics[width=30mm]{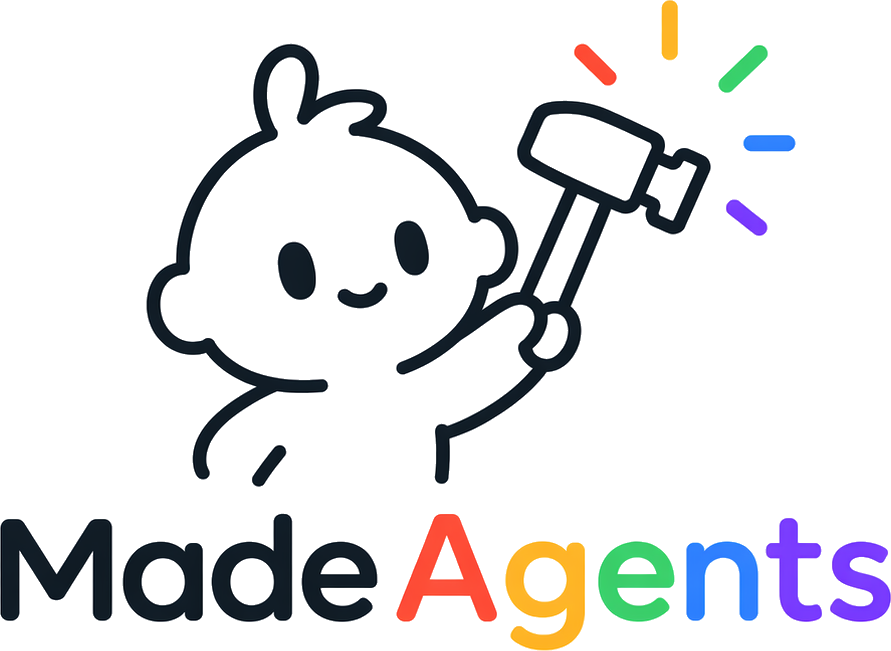}}
    }
\end{abox}

\vskip 8mm

\input{sections/introduction}

\input{sections/concept}

\input{sections/agent}

\input{sections/multi_device}

\input{sections/multi_agent}

\input{sections/security}

\input{sections/deployment}

\input{sections/conclusion}

\vskip 8mm
\bibliography{references}
\bibliographystyle{unsrt}

\end{document}

%% file: sections/introduction.tex
\section{Introduction}

Large language models (LLMs) have evolved AI assistants from conversational interfaces into autonomous reasoning engines~\cite{wei2022cot,yao2023react}. Supported by various open-source agent frameworks~\cite{autogpt2023,chase2022langchain,xie2023openagents,wu2023autogen}, assistants are now expected to maintain context, invoke tools, and pursue complex tasks over time. In this paradigm, an agent is no longer merely a chatbot, but a computational actor deeply integrated with its digital environment.

\begin{figure}[htbp]
    \centering
    \includegraphics[width=0.8\textwidth]{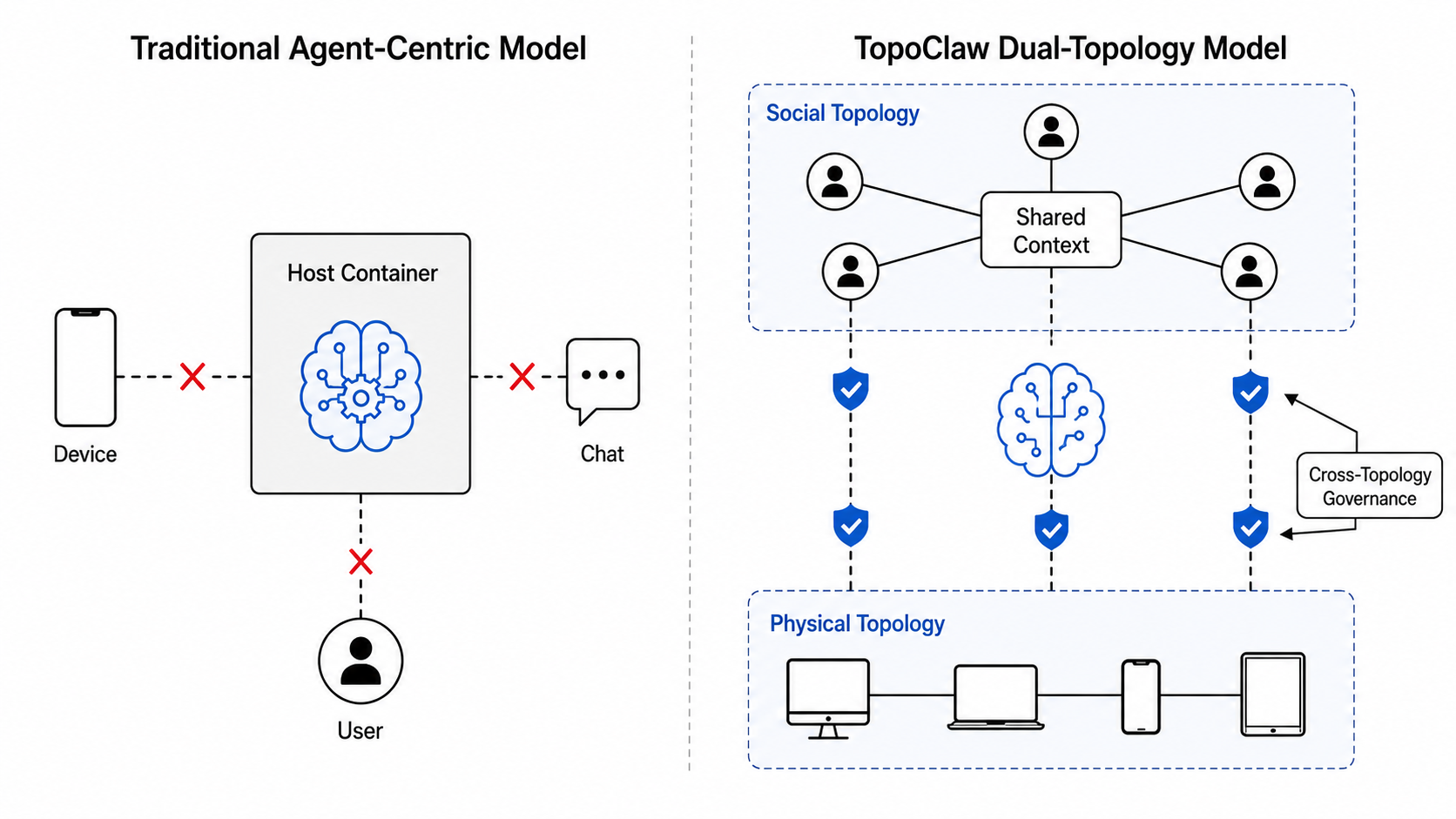}
    \caption{Paradigm shift from traditional agent-centric isolation to TopoClaw's human-centric dual-topology model. The agent is liberated from a single host container (left) and acts as a digital twin navigating both the user's physical device network and social collaboration spaces, secured by cross-topology governance (right).}
    \label{fig:motivation}
\end{figure}

This shift has motivated the development of the Agent Operating System (Agent OS). Early literature frames the Agent OS as a kernel-like layer responsible for agent lifecycle management—handling memory, scheduling, and access control~\cite{mei2024aios,zhao2024aiosurvey}. Practical Agent OS implementations further demonstrate how persistent agents can be brought into daily use through local execution and extensible capabilities~\cite{openclaw2026,hermes2026,genericagent2026}. In parallel, work on GUI agents shows that models can navigate concrete interfaces like phones and desktops~\cite{wen2024autodroid,ye2025mobileagentv3,hu2025osagents}. Together, these efforts establish the essential infrastructure to keep agents running persistently and reliably.

Despite this progress, the dominant Agent OS design remains largely agent-centric. The OS is primarily viewed as a runtime container that confines the agent to a single host machine, focusing on sustaining its internal processes (reasoning, memory, tool invocation). While this foundation is essential, it leaves a critical gap: it fails to address how these actions should integrate into the real-world workflows of the human user. To truly serve as a daily assistant, an Agent OS must shift from an agent-centric container to a human-centric execution environment. It must release the agent from the confines of a single machine and allow it to navigate the complex, distributed topology of the user's actual digital life.

This transition is necessary because the human's digital environment is not a flat, isolated workspace, but rather a complex structure defined by two fundamental topologies:

First, humans operate within a Physical Device Topology. A user's context is inherently fragmented across multiple heterogeneous devices (e.g., desktops, smartphones, edge sensors), each with asymmetric hardware capabilities and distinct software ecosystems. When an agent is confined to a single node, it cannot follow the user's physical trajectory. A human-centric OS must treat this fragmented hardware landscape as a unified execution plane, allowing the agent's cognition to decouple from physical actuation and route actions across the device topology.

Second, human workflows are deeply embedded in a Social Relationship Topology. Tasks frequently occur in shared communication spaces involving multiple users, teams, and other agents. Traditional agents act as isolated tools; however, when an agent enters a collaborative environment—such as negotiating a meeting in a group chat—it must act as a socially-aware entity. Its actions must remain explicitly tied to the identity, role, and permissions of its human owner. Without this attribution, multi-agent collaboration collapses into untraceable automation.

Finally, navigating these dual topologies introduces unprecedented challenges for accountable autonomy. As agents cross physical device boundaries and interact within social networks, traditional static permissions and post-hoc confirmation dialogs fail. Users need a dynamic, context-aware governance mechanism that enforces safety boundaries across both the physical and social topologies, ensuring the agent operates freely but strictly within its delegated authority.

Existing Agent OSs provide limited support for managing this distributed, social, and sensitive reality, often offloading these challenges to ad hoc application integrations. TopoClaw addresses this gap by presenting a human-centered, topology-aware Agent OS. By modeling the human's digital ecosystem as interconnected physical and social networks, TopoClaw unifies device operation, messaging, and skills around a single question: how should agents act across devices, social contexts, and authority boundaries on behalf of users?

The contributions of TopoClaw are threefold:

\begin{itemize}
    \item \textbf{Cross-Device Action Placement.} To navigate the physical topology, TopoClaw models the user's heterogeneous devices as a unified execution cluster. It decouples intent generation from physical execution, shifting scheduling from isolated tool calls to distributed actions that are dynamically routed and placed across the device network based on hardware affordances and user context.
    \item \textbf{Cross-User Identity Attribution.} To navigate the social topology, TopoClaw treats agents as "Digital Twins" within a multi-user environment. It enables agents owned by different users to coordinate in shared collaborative spaces while strictly preserving identity provenance, role-aware permissions, and an unbroken chain of human accountability.
    \item \textbf{Cross-Context Authority Governance.} To secure these dual topologies, TopoClaw pairs broad operational capability with a distributed, context-aware policy enforcement architecture. It intercepts and evaluates actions across both the physical device boundaries and the social trust boundaries, turning safety from a localized prompt into a structural, OS-level guarantee that enables safe proactive autonomy.
\end{itemize}

This report presents TopoClaw as an engineering-oriented reference architecture for a human-centered, topology-aware Agent OS. The following sections describe its design principles, runtime architecture, cross-device execution model, collaboration mechanisms, security architecture, and deployment model.

%% file: sections/concept.tex
\section{Motivation \& Design Principles}

\subsection{The Limits of Isolation}

The emergence of LLM-driven agents has profoundly expanded the possibility space of AI assistants, transitioning them from passive text generators to active planners and tool users. This paradigm shift has given rise to the concept of the Agent Operating System (Agent OS). However, when examining early literature and frameworks in this domain (e.g., AIOS~\cite{mei2024aios}, AutoGPT~\cite{autogpt2023}), a critical pattern emerges: existing Agent OS designs are overwhelmingly agent-centric. 

In the agent-centric paradigm, the "Operating System" is conceptualized primarily as a lifecycle manager for the model. Its primary concerns are maintaining the agent's internal state, keeping memory vectors intact, scheduling LLM inferences, and orchestrating tool invocations. While this foundation is undeniably necessary, it operates under a limiting assumption: it treats the agent's execution as a closed, isolated process, detached from the complexities of the user's broader digital ecosystem. 

When deployed into real-world scenarios, the limitations of this isolated model become immediately apparent. The human digital environment is not a flat, single-device workspace, but rather a complex structure defined by two fundamental topologies. First, humans operate within a Physical Device Topology. Information and hardware capabilities are fragmented across personal computers, edge sensors, and mobile devices. An isolated agent confined to a single host cannot follow the user's physical trajectory or leverage asymmetric hardware capabilities. Second, human workflows are embedded in a Social Relationship Topology. Tasks occur in shared spaces involving multiple users and teams. An isolated agent acting as an anonymous script cannot participate in these collaborative environments without causing ambiguous accountability~\cite{park2023generative,hong2023metagpt}.

Consequently, navigating these dual topologies introduces a severe governance challenge. As agents move from observation to operation—executing shell commands or communicating externally—traditional static permissions and post-hoc approval dialogs fail~\cite{saltzer1975protection,nist2020zerotrust}. Users cannot be expected to manually approve every micro-step of a background task across different devices and chats, yet they cannot blindly trust a black-box model operating outside a strict trust boundary.

\subsection{A Topology-Aware Architecture}

To build an assistant capable of genuine utility, the architecture must invert its focus: the OS must not be built around the agent's internal processes, but around the human's external environment. Rather than forcing the human to adapt to an agent-centric execution model, TopoClaw restructures the OS to adapt to the human. This paradigm is realized through three core design principles that directly address the physical, social, and governance challenges:

\textbf{Distributed Action Placement.} To navigate the physical topology, the OS must collapse the boundaries between devices. TopoClaw establishes a unified execution plane where cognition is decoupled from physical actuation. Actions and context flow seamlessly with the user, dynamically routed to the most appropriate node in the hardware cluster.

\textbf{Identity-Preserving Delegation.} To navigate the social topology, the OS must introduce a native identity layer. When an agent operates in shared human-agent spaces or spawns sub-agents, it must act as a Digital Twin. It must strictly maintain the provenance, role, and accountability of its human delegator, ensuring that multi-agent collaboration remains traceable.

\textbf{Cross-Topology Boundary Defense.} To enforce bounded delegation without crippling autonomy, safety must be elevated from application-level confirmation dialogs to cross-topology, OS-level capability controls~\cite{owasp2025llmtop10}. By embedding layered authorization directly into the runtime, TopoClaw guarantees that all actions—whether traversing physical devices or operating in shared social groups—remain strictly governed by the user's explicit boundaries.

By adopting this topology-aware approach, TopoClaw delivers an Agent OS where multi-device routing, multi-user identity, and cross-context security are not optional plugins, but the very foundation of the architecture.

%% file: sections/agent.tex
\section{Architecture Overview}

The preceding section argued that a human-centric Agent OS must elevate physical fragmentation, social delegation, and bounded autonomy into runtime primitives. This section answers the complementary question: what shape of architecture can carry those primitives end-to-end? TopoClaw's answer is an event-driven runtime in which cognition, actuation, collaboration, and policy enforcement are decoupled yet jointly observable. The agent is no longer a single synchronous loop confined to one host; it becomes a distributed actor whose steps are placed, attributed, and governed as they cross topological boundaries.

\subsection{A Decoupled Runtime}

Conventional agent stacks often collapse planning, tool execution, messaging, and permission checks into one process boundary. That design matches an agent-centric mental model, but it breaks down when actions must traverse the user's physical and social topologies. TopoClaw therefore replaces the monolithic loop with a decomposable runtime connected by an asynchronous message bus~\cite{hong2023metagpt}. Heavy reasoning, lightweight edge control, outbound collaboration, and authorization can progress at different rates and on different nodes without losing a single traceable chain of custody.

\begin{figure}[t]
    \centering
    \includegraphics[width=\textwidth]{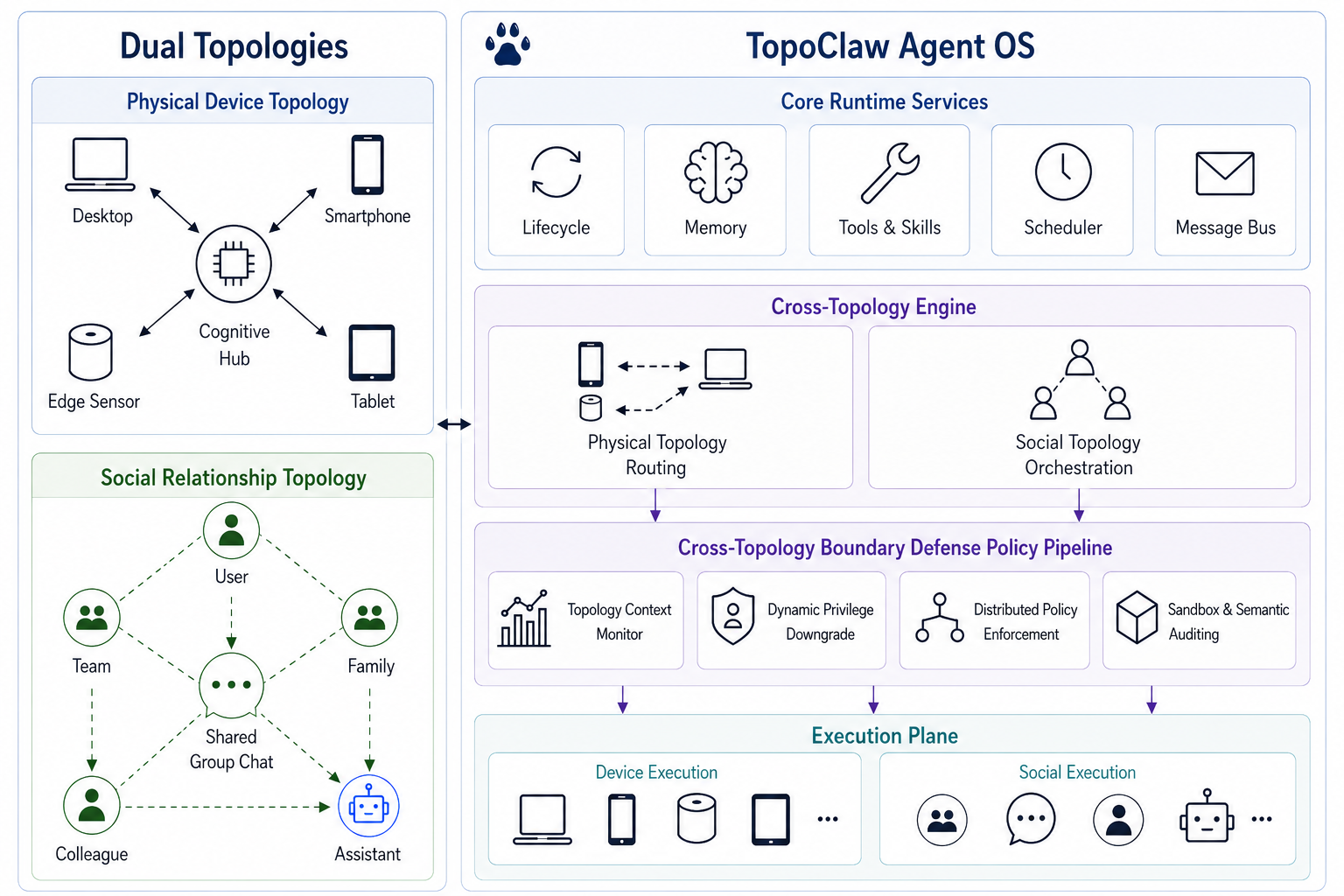}
    \caption{TopoClaw System Architecture. The OS acts as a decoupled runtime navigating dual topologies. Intent generation (physical routing and social orchestration) is decoupled from the execution plane, with all actions strictly intercepted by the cross-topology boundary defense pipeline.}
    \label{fig:architecture}
\end{figure}

Figure~\ref{fig:architecture} summarizes TopoClaw as a cooperating set of runtime mechanisms designed to navigate the human environment. These mechanisms provide the structural foundation for the OS:

\paragraph{Physical Topology Routing.}
To address hardware fragmentation, the runtime treats heterogeneous devices and durable workspace state as a single logical execution plane. The cognitive hub decides what must happen, while the placement engine and synchronization protocols decide where it runs and how artifacts cross spatial gaps. Furthermore, temporal continuity is maintained through layered, inspectable plain-text memory, ensuring that context remains an OS concern the human can audit~\cite{packer2023memgpt}.

\paragraph{Social Topology Orchestration.}
To address social delegation, the runtime sits above raw execution and treats every outward-facing step as an attributed event. Ownership, role, and delegation scope are carried alongside messages and spawned workflows. This ensures that when agents participate in shared channels or delegate to sub-agents, multi-agent graphs can scale without severing the link between the automated action and the human delegator.

\paragraph{Cross-Topology Boundary Defense.}
Broad capability across devices and groups increases the blast radius of potential errors. The governance mechanism wraps the event bus and node boundaries so that resource access, command execution, and sensitive side effects pass through layered, context-aware policy enforcement. Safety is thus structural: proactive and background work can run only where the runtime can prove compliance with the user's declared boundaries.

The architecture overview is intentionally operational: it states why the runtime is partitioned and which problems each partition solves. The remainder of this report details the formal mechanisms behind these partitions. Section~\ref{sec:multi_device} details physical topology navigation and action placement; Section~\ref{sec:multi_agent} details social topology orchestration and identity provenance; Section~\ref{sec:security} details the boundary defense model and safe proactivity. The deployment section then closes the loop on how this reference architecture is packaged for daily use.

%% file: sections/multi_device.tex
\section{Navigating the Physical Topology}
\label{sec:multi_device}

The human digital environment is fundamentally fragmented. In daily life, tasks are scattered across a physical topology of different devices, operating systems, and edge sensors~\cite{hu2025osagents}. Traditional AI assistants are confined to a single host machine, unable to follow the user's physical trajectory. TopoClaw addresses this limitation by modeling the user's hardware ecosystem as a unified physical topology, allowing the agent to decouple cognition from actuation and route actions across device boundaries.

\subsection{The Physical Device Graph}

To achieve human-centric spatial continuity, TopoClaw treats the user's paired devices as a physical topology graph $\mathcal{G}_{phys} = (\mathcal{N}, \mathcal{E}_{sync})$. Here, $\mathcal{N} = \{n_1, n_2, \dots, n_k\}$ represents the set of available execution nodes (e.g., a desktop PC, a smartphone), each characterized by a distinct hardware and software capability profile $\mathcal{C}(n_i)$. The edges $\mathcal{E}_{sync}$ represent the OS-level state synchronization bus connecting them.

When a user issues a complex command, the central cognitive hub generates a logical intent, decomposing the workflow into a directed acyclic graph (DAG) of actions $G_{task} = (V, E)$. The placement engine then solves a graph mapping problem: it computes a routing function $f: V \rightarrow \mathcal{N}$ such that the required affordances of each action $v \in V$ are optimally satisfied by the target node's capabilities $\mathcal{C}(f(v))$.

For example, a task requiring deep filesystem retrieval and subsequent SMS communication is seamlessly mapped across the topology: the retrieval node $v_{read}$ is routed to the desktop, the payload transfers across the synchronization edge, and the dispatch node $v_{sms}$ is routed to the mobile edge node. This spatial decoupling ensures that the agent's operational reach matches the user's distributed footprint, allowing context to flow automatically without the human acting as an intermediary router.

\subsection{Temporal State Management}

Just as execution must span the physical device graph, the agent's context must span multiple sessions over time. A genuinely human-centric OS must ensure that this global "memory" remains transparent and controllable by the user, rather than trapped in the opaque embeddings of a vector database.

TopoClaw maintains temporal continuity by treating memory as the global state of the physical topology~\cite{packer2023memgpt}. Formally, this temporal context is maintained as a human-readable state tuple $\mathcal{S} = \langle M_{short}, M_{long}, M_{log} \rangle$, representing the short-horizon focus buffer, the long-term knowledge base, and the append-only chronological audit log, respectively. An asynchronous background consolidation function $\Phi$ updates this state after each interaction: $\mathcal{S}_{t+1} = \Phi(\mathcal{S}_t, \text{Observation}_t)$.

By storing this state tuple entirely in structured, plain-text formats within the user's workspace, TopoClaw returns the ultimate authority over the agent's knowledge to the human operator. The temporal context flows with the user across the physical topology, but because it relies on inspectable text, the user always holds the editorial key to correct or modify the agent's state directly.

%% file: sections/multi_agent.tex
\section{Navigating the Social Topology}
\label{sec:multi_agent}

Human workflows are deeply embedded in social relationships. Tasks frequently occur in shared communication spaces involving multiple users, teams, and other agents. However, traditional agents are designed as isolated tools; when they enter a collaborative environment, they operate as anonymous automation scripts, obscuring accountability. To integrate into real-world workflows, TopoClaw models the multi-user environment as a social topology, ensuring agents act as attributed participants.

\subsection{Digital Twins as Nodes}

In TopoClaw, the collaborative ecosystem is modeled as a social topology graph $\mathcal{G}_{soc} = (\mathcal{U}, \mathcal{E}_{trust})$, where nodes $\mathcal{U}$ represent human users and edges $\mathcal{E}_{trust}$ represent communication channels and trust delegations. An agent does not exist outside this graph; it enters the social topology as a Digital Twin of its human owner~\cite{kritzinger2018digitaltwin}.

To maintain this relationship, any action or message $m$ traversing a social edge is never transmitted in isolation. The OS wraps it into an attributed event tuple $E = \langle m, \mathcal{I}_{human}, \mathcal{I}_{twin}, \rho \rangle$, where $\mathcal{I}_{human}$ represents the verifiable identity of the human delegator, $\mathcal{I}_{twin}$ is the specific agent instance, and $\rho$ denotes the delegated role.

When an agent responds in a shared group, this tuple ensures that the social accountability for the agent's words and actions rests firmly with its human delegator. The agent inherits the user's social graph, allowing it to act as a seamless proxy in human-agent interactions without breaking the chain of provenance.

\subsection{Shared Context Spaces}

This strict provenance mechanism forms the foundation for safe multi-agent interoperability~\cite{wu2023autogen}. In the social topology, group chats and collaborative projects act as shared context spaces where multiple nodes intersect. 

TopoClaw supports advanced operational modes where messages within these spaces are broadcasted to all participating Digital Twins. Because each agent operates based on its owner's unique data context and maintains transparent social attribution via the event tuple $E$, they can engage in complex intent exchange and task negotiation. For instance, two agents can autonomously negotiate a meeting time on behalf of their respective owners within a shared chat. The entire negotiation process remains fully visible, traversing the social edges while preserving an unbroken chain of human accountability. By elevating identity to a runtime primitive, TopoClaw ensures that multi-agent graphs can scale organically in social environments.

%% file: sections/security.tex
\section{Cross-Topology Governance}
\label{sec:security}

The ultimate test of a human-centric Agent OS is how it handles autonomy. As agents navigate the physical device topology and interact within the social relationship topology, the traditional security model of static permissions and post-hoc approval dialogs breaks down~\cite{nist2020zerotrust,owasp2025llmtop10}. A truly autonomous agent must perform background processing and proactive tasks, requiring a dynamic governance mechanism that defends the boundaries of both topologies.

\subsection{Defending Boundaries}

Because an agent's actions span physical devices and social networks, the context of any execution request is highly dynamic. An action might originate from the user's desktop, a scheduled cron job, or a command issued by a colleague across a social edge. TopoClaw replaces static permissions with a distributed, context-aware policy enforcement architecture designed to defend these topological boundaries~\cite{saltzer1975protection}.

\textbf{1. Physical Boundary Defense (Distributed PEPs).} Actions routed across the physical topology graph $\mathcal{G}_{phys}$ cannot rely solely on the central hub for security. TopoClaw utilizes Distributed Policy Enforcement Points (PEPs) at the edge nodes. The safety of an action $a$ under a dynamic context $c$ is determined by a boolean conjunction across all $k$ enforcement layers:
\begin{equation}
\text{Safe}(a, c) = \bigwedge_{i=1}^{k} \Pi_i(a, c)
\end{equation}
where $\Pi_i$ represents the policy evaluation function at layer $i$. Even if the cognitive hub is compromised, the receiving edge device independently verifies the action against local boundaries before actuation.

\textbf{2. Social Boundary Defense (Dynamic Privilege Downgrade).} When an agent receives a request across an edge in the social topology graph $\mathcal{G}_{soc}$ (e.g., from a colleague in a shared chat), TopoClaw implements strict permission inheritance to prevent privilege escalation. Let $\mathcal{P}_{baseline}$ denote the agent's maximum authorized privilege set granted by its owner, and $\mathcal{P}_{req}$ represent the privilege set of the external requesting entity. The effective execution privilege $\mathcal{P}_{eff}$ is strictly bounded by their intersection:
\begin{equation}
\mathcal{P}_{eff} = \mathcal{P}_{baseline} \cap \mathcal{P}_{req}
\end{equation}
This mathematical confinement defends the social boundary, ensuring that an external participant cannot trick the agent into accessing the owner's private resources.

\textbf{3. Structural Sandbox Confinement.} To defend the boundary between the agent and the host OS, TopoClaw restricts all write operations to a configured workspace scope. Any attempt to modify resources outside this structural sandbox triggers a strict halt, bounding the blast radius of background tasks.

\textbf{4. Semantic Command Auditing.} At the execution layer, TopoClaw performs semantic auditing of pending operations, unconditionally intercepting inherently dangerous idioms and unauthorized privilege escalations before they reach the host kernel.

\subsection{Safe Proactivity}

It is precisely because these topological boundaries are mathematically and structurally defended that TopoClaw can safely introduce advanced autonomy features. 

Rather than passively waiting for user input, TopoClaw's context-aware proactivity loop periodically wakes the agent to inspect pending tasks or monitor system events. Because these unprompted background invocations are routed through the exact same rigorous boundary defense pipeline as direct user commands, proactive actions are intrinsically safe. By making cross-topology governance a foundational OS-level guarantee, TopoClaw transforms proactive execution from a security risk into a reliable mechanism for daily human workflows.

%% file: sections/deployment.tex
\section{Implementation and Ecosystem}

To validate the theoretical architecture proposed in previous sections, we implemented TopoClaw as a fully functional, open-source Agent OS. The ecosystem extends beyond the core routing and governance engines to encompass a dynamic capability injection mechanism and a template registry, allowing the system to scale its utility without modifying the underlying OS kernel.

\subsection{Dynamic Capability Injection}

In the physical topology, while the hardware of edge nodes is fixed, their software affordances must be dynamically extensible. TopoClaw implements a distributed capability registry (the Skill Ecosystem) that allows the OS to inject new tools into the execution plane at runtime~\cite{anthropic2024mcp,xie2023openagents}.

The capability injection mechanism is multi-sourced, supporting official public registries, community-driven hubs, and user-generated workspaces. Rather than hardcoding tool logic into the agent's prompt, skills are packaged as standardized manifests that declare their required dependencies, target execution environment (e.g., PC or Mobile), and semantic descriptions. When a skill is invoked, the placement engine automatically routes the execution to the node that satisfies the skill's environmental constraints.

Beyond external registries, TopoClaw ships with a baseline inventory of 22 built-in capabilities designed to cover fundamental cross-topology scenarios out of the box. These built-in capabilities span four functional categories:
\begin{itemize}
    \item \textbf{Utility Capabilities:} Core functions for information retrieval, file management, and cron-based scheduling.
    \item \textbf{Cross-Device Capabilities:} Edge-specific functions such as mobile gallery management, clipboard synchronization, and deep link invocation.
    \item \textbf{Social Capabilities:} Mechanisms for group messaging, digital twin identity assertion, and contact management.
    \item \textbf{System Capabilities:} OS-level tools for managing the layered memory state and authoring new skills.
\end{itemize}

\subsection{Topology Template Registry}

While skills provide atomic capabilities, TopoClaw also supports the serialization and distribution of complex agent personas. We implemented the Assistant Plaza as a Topology Template Registry. 

A configured assistant—comprising its specific system prompt, bounded skill set, and behavioral defaults—is essentially a pre-configured "Digital Twin Template." The registry allows users to package these templates and distribute them across the social topology. The design philosophy of this registry rests on three pillars:
\begin{itemize}
    \item \textbf{Accessibility:} Any user can serialize their tuned assistant into a portable template and publish it to the registry.
    \item \textbf{Transparent Provenance:} Each published template carries explicit metadata regarding its author and intended use case, maintaining trust within the social topology.
    \item \textbf{Environment Adaptation:} Templates are published with specific platform constraints. Upon instantiation, the OS automatically binds the assistant to the appropriate execution node in the physical topology without requiring manual reconfiguration.
\end{itemize}

Furthermore, TopoClaw enables direct peer-to-peer sharing of these templates. A template can be serialized into a rich-format message and transmitted across a social edge (e.g., sent to a colleague in a chat). Upon receipt, the OS instantly instantiates the template, allowing users to rapidly propagate specialized workflows across their collaborative networks.

\subsection{Deployment Configurations}

TopoClaw is engineered to be highly modular, supporting three distinct deployment configurations that instantiate different scales of the dual-topology model based on user requirements:

\textbf{1. Single-Node Topology (Desktop Standalone).} In this lightweight configuration, the physical topology collapses to a single node (the PC). The cognitive hub and execution engine run locally, providing a robust baseline for individual users who require an isolated, privacy-first assistant without cross-device routing.

\textbf{2. Social-Only Topology (Cross-User Collaboration).} This configuration introduces the message routing backend to the standalone mode. It activates the social topology, enabling state synchronization and identity provenance between multiple users. This is optimized for team collaboration scenarios where agents must operate as attributed digital twins within shared group chats.

\textbf{3. Full Dual-Topology (Cross-Device Complete).} The complete instantiation of the TopoClaw architecture. It integrates the central cognitive hub, the message routing backend, and mobile edge nodes. This configuration fully activates both the physical and social topologies, enabling the agent to route actions across heterogeneous devices while maintaining strict identity and governance boundaries in multi-user environments.

%% file: sections/conclusion.tex
\section{Conclusion}

The transition from isolated conversational models to autonomous, daily-use AI assistants requires a fundamental shift in operating system design. Traditional Agent OS architectures remain agent-centric, confining execution to a single host and treating the human's complex digital environment as an afterthought. 

This report presented TopoClaw, an open-source, human-centric Agent OS designed to invert this paradigm. By modeling the human digital environment as a dual-topology structure—comprising a Physical Device Topology and a Social Relationship Topology—TopoClaw elevates real-world environmental constraints into first-class runtime primitives. 

Our architecture demonstrates three core contributions:
\begin{itemize}
    \item \textbf{Distributed Action Placement:} By decoupling cognition from physical actuation, TopoClaw enables actions to be dynamically routed across a heterogeneous device cluster, solving the challenge of hardware fragmentation.
    \item \textbf{Social Orchestration:} By treating agents as attributed "Digital Twins," the OS ensures that multi-agent collaboration in shared spaces maintains strict identity provenance and an unbroken chain of human accountability.
    \item \textbf{Cross-Topology Governance:} By implementing distributed, context-aware policy enforcement at the boundaries of both physical and social topologies, TopoClaw transforms safety from a static, localized prompt into a structural guarantee, enabling safe proactive autonomy.
\end{itemize}

Supported by a robust ecosystem for dynamic capability injection and topology template sharing, TopoClaw proves that multi-device routing, multi-user identity, and cross-context security need not be brittle, application-level patches. Instead, when embedded deeply into the OS runtime, they unlock a new class of human-centric assistants capable of seamlessly navigating the distributed, social, and sensitive realities of daily human workflows.